# From Intra-Datacenter Interconnects to Metro Networks: Does CV-QKD Need Loss- or Bandwidth-Conscious Receivers?

Florian Honz[1], Fabian Laudenbach[2], Hannes Hübel[1], Philip Walther[3], and Bernhard Schrenk[1]

[1] AIT Austrian Institute of Technology, 1210 Vienna, Austria. florian.honz@ait.ac.at
[2] AIT Austrian Institute of Technology, 1210 Vienna, Austria (now with Xanadu Quantum Tech).
[3] University of Vienna, Faculty of Physics, 1090 Vienna, Austria.

**Abstract** *We experimentally compare a loss-optimized coherent heterodyne and a bandwidth-blessed intradyne CV-QKD architecture. We find the former to prevail performance-wise for medium/long link reach, while the latter features a 5-9 dB higher secure-key rate over short reach.* ©2022 The Author(s)

**Introduction**
The proper protection of sensible data is an ever-increasing demand of our ICT-based society. With the rapid progress of quantum computers, the current cryptographic key exchange methods based on public key infrastructure are no longer considered safe. A particular attractive version is the so-called continuous variable (CV) QKD, mostly due to the overlap with modulation and detection techniques as found in coherent optical communication systems. The implementation of CV-QKD therefore profits from the tremendous advancement in optical components and, more importantly, does not require bulky single photon detectors, thus opening a path for a full QKD system to be integrated on photonic chips that could even be co-packaged with classical systems. Various CV-QKD systems have been demonstrated, including co-transmitted local oscillators (LO) [1,2], and receiver-side LO schemes [3-10] that involved LO training due to the weak quantum signal that is received at a low SNR < 1. Since the performance of CV-QKD is very much determined by loss and noise, the receiver is the crucial part of a CV-QKD system [9]. Moreover, the highly varying optical budgets in various telecom and datacom networks call into question whether a single CV-QKD receiver architecture can address all metro, access and datacenter requirements.

In this work we address this question through an experimental study of two CV-QKD architectures building on (*i*) a bandwidth-optimized phase-diversity (coherent intradyne) detection with simultaneous measurement of both quadratures of the quantum signal, and (*ii*) a simplified CV receiver employing a low-loss 180° hybrid in a coherent heterodyne detection scheme to avoid spectral folding of the quantum signal for digitized in-phase/quadrature (I/Q) recovery at the expense of precious receiver bandwidth. We will experimentally evaluate the supported secure-key rates (SKR) over the link reach and show that there is no one-fits-all architecture but an application-dependent need for loss/bandwidth consciousness.

**Loss- and Bandwidth-Conscious CV-QKD**
Figure 1 presents the two receiver architectures in the overall experimental configuration under a CV transmission methodology that employs a polarization-multiplexed pilot tone for the purpose of optical frequency and carrier-phase recovery. Baseline to both schemes is a quantum receiver with a clearance-limited bandwidth $B_q$. Exploiting its full bandwidth aiming at highest-possible secure key rates would at first glance point towards coherent intradyne reception, where the optical frequencies of quantum signal and LO are very similar, yielding a quantum heterodyne measurement of the $I,Q$ quadratures through an optical 90° hybrid. However, the loss of this optical mixer for LO and quantum signal is a critical parameter for CV-QKD, thus motivating a more loss-conscious receiver architecture. This is found with coherent heterodyne detection, which trades optical phase-diversity reception with a quantum homodyne measurement using a simpler 180° hybrid. If the optical quantum signal is down-converted to an electrical intermediate frequency (IF) that adheres to the condition IF > $R_q$, no spectral folding occurs for the detected signal and thus the full information on its complex-valued optical amplitude is preserved, allowing for the restoration of amplitude and phase. However, off-loading the function of the optical 90° hybrid to the digital domain implies that the quantum receiver has to cope with a high IF frequency where a response roll-off and increased TIA noise might apply.

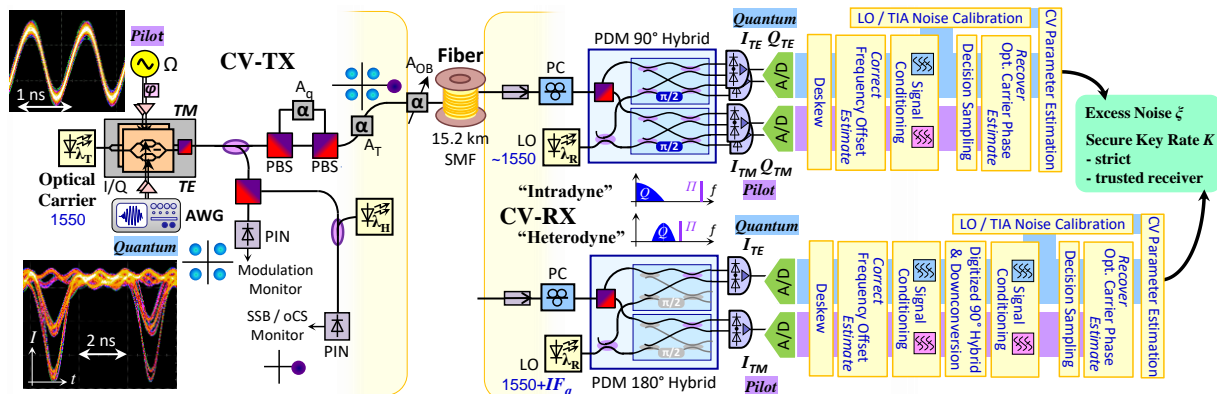

**Fig. 1:** Experimental setup & DSP stacks for comparing coherent intradyne and heterodyne CV receiver performances

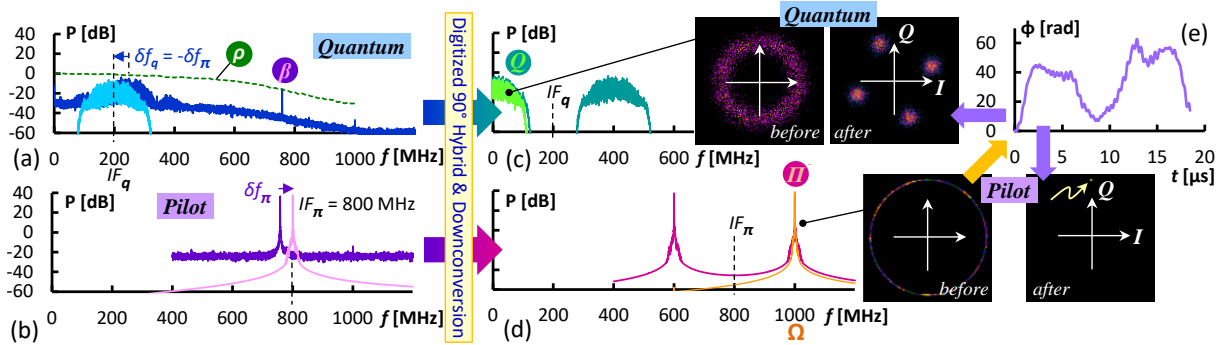

**Fig. 2:** CV reception through coherent heterodyne system in (a, c) quantum and (b, d) pilot plane. (e) Estimated carrier phase.

**Experimental CV-QKD Evaluation**
The CV transmitter builds on the joint transmission of a discretely modulated quantum signal with a polarization-multiplexed pilot tone. Optical I/Q modulation imprints the QPSK-encoded quantum signal at a (variable) symbol rate $R_q$. The optical carrier at 1550 nm is further modulated with a carrier-suppressed (oCS) single-sideband (SSB) pilot tone at $\Omega$ = 1 GHz. This modulation scheme reduces in-band pilot crosstalk noise to the quantum signal (oCS) and the pilot (SSB) during coherent reception. A polarization-selective attenuator sets a power ratio of 23.4 dB ($A_q$) between pilot and quantum signal to ensure a high SNR for pilot reception. The quantum signal is transmitted with a power of 4 photons/symbol ($A_T$). This CV signal is then sent to either CV receiver through 15.2 km of ITU-T G.652B compatible single-mode fiber (SMF). The optical budget of the link was varied ($A_{OB}$) to evaluate the QKD performance.

At the CV receivers a LO with a power of 13 dBm beats with the received signal in a polarization-diversity optical mixer. Noise- and bandwidth-optimized balanced detectors are then dedicated to the quantum and pilot polarization planes. The respective detectors had bandwidths of $B_q$ = 360 MHz and 1.6 GHz and featured a CMRR of 47.3 and 38 dB, respectively. The clearance for the quantum receiver was >20 dB for an LO power of 10 mW.

For the *bandwidth-conscious intradyne CV receiver* the 90° hybrid as the optical mixer directly yields the $I,Q$ quadratures, allowing us to place the quantum signal with a wider $R_q$ = 250 Mbaud at the baseband. The LO is matched to the optical carrier wavelength at the transmitter and the involved digital signal processing (DSP) uses pilot-based disciplining to recover the optical frequency offset and the phase evolution Φ for the optical carrier-phase correction of the quantum signal. Noise calibration is performed on the LO noise and the TIA noise of the balanced detectors and used for the purpose of CV parameter estimation.

For the *loss-conscious heterodyne CV receiver*, we used a 180° hybrid combined with only one balanced receiver for each plane. A detuning of $IF_q = 2R_q$ for the LO, at $R_q$ = 100 Mbaud, enables us to recover the $I,Q$ quadratures in the digital domain through DSP-based mixing of pilot and quantum signal with a complex carrier in a digitized 90° hybrid. Figure 2 presents the corresponding signal spectra. For the sake of visual clarity, the signals are plotted for a launch power of ~400 photons/symbol.

The received signals in the quantum and pilot polarization planes are reported in Fig. 2a/c and 2b/d, respectively. The LO detuning by $IF_q$ places the quantum signal between $R_q$ = 100 MHz and the roll-off ρ of the quantum receiver (Fig. 2a), with the pilot tone falling at $IF_\pi$ = 800 MHz < Ω (Fig. 2b). Its bleed-through β (Fig. 2a) to the quantum plane is attributed to the finite polarization extinction of the optical mixer. The residual frequency offset $δf_\pi$ (Fig. 2b) can be determined with a high SNR and compensated. The signals are then mixed with a complex carrier with $f = IF_q$ to yield the quadratures for the quantum signal ($Q$ in Fig. 2c) and the pilot tone ($\Pi$ in Fig. 2d) at the baseband and at $\Omega$ = 1 GHz, respectively. Spectral mirror components can be easily filtered. Information on the phase evolution $\Phi$ (Fig. 2e) of the optical carrier can be obtained from the recovered pilot, again at a high SNR, and used for carrier-phase recovery of the quantum signal. The typical peak phase velocity was ~35 rad/µs.

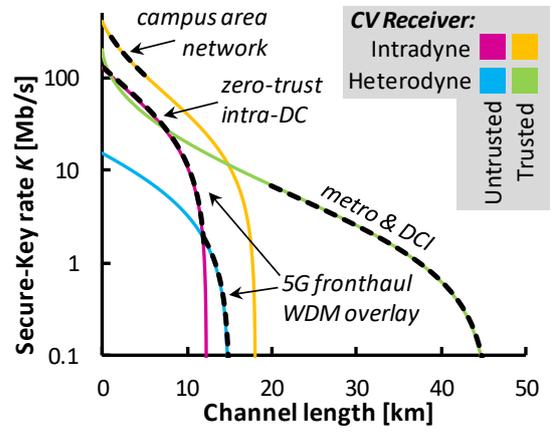

**Fig. 3:** CV-QKD performance estimation of the secure-key rate based on the experiment.

**CV-QKD Performance and Applicability**

The excess noise parameter $\xi$, as the quadrature variance in addition to the obligatory quantum shot noise, is used as a primary performance indicator since every CV-QKD system is sensitive to noise, regardless of the chosen protocol. A strict untrusted receiver noise model and a trusted model, where the receiver is placed at a trusted location, were considered. These take into account the total excess noise $\xi$ and, alternatively, the excess noise $\xi_T$ excluding the TIA noise. For example, the heterodyne CV receiver yields $\xi$ = 0.0182 SNU and $\xi_T$ = 0.0105 SNU when referred to the receiver input. Offline simulations based on the experimentally gathered data for different channel loss conditions ($A_{OB}$ + 15.2 km of SMF) were performed to derive an estimate for the supported SKR, taking into consideration the channel loss, the total detection loss and the excess noise $\xi$, $\xi_T$. In addition, a reconciliation efficiency of 0.97 was assumed. The results are reported in Fig. 3. For a SKR of 1 Mb/s, we can reach a transmission reach of 13.2 and 12.1 km in case of an untrusted heterodyne and intradyne receiver, respectively. This link reach extends to 37.4 and 17.8 km under a trusted receiver noise model. For the trusted heterodyne receiver, a SKR of 10 Mb/s can be supported over 16.2 km.

These performances can be translated to network applications as follows: Zero-trust intra-datacenter links (0 – 10 km) require untrusted receivers, which renders the bandwidth-conscious intradyne scheme as favorable. This domain promises a SKR of more than 10 Mb/s. To estimate the total data capacity that can be secured through this SKR, we apply the NIST recommendation for AES key renewal. In this case, a single 256-bit long AES key should only be used for encrypting a maximum of 64 Gbyte of data. With the given SKR equivalent to a generation rate of 39000 AES keys/s, this means that a capacity of ~20000 Tb/s could be secured. Thus, a single CV-QKD channel can protect a massive multi-Pb/s optical interconnect. It is important to note here, that the estimated key-rates were derived for dark fibers and Gaussian modulation, though discrete modulation was used in the present experiment.

In case of 5G fronthaul links (5 – 20 km) that are conceived as virtual p2p WDM overlays, the untrusted receiver model favors again the intradyne scheme, up to an inflection point of 11.9 km beyond which the loss-conscious heterodyne receiver becomes more favorable. For applications such as campus area networks (1 – 5 km) or metro links and datacenter interconnects (20 – 50 km), a trusted receiver hosted at a secured location can be considered. In such a scenario, bandwidth-conscious intradyne receivers promise a SKR of >100 Mb/s for short-reach campus networks, while longer-reach metro/DCI clearly require loss-conscious heterodyne receivers for which a SKR of >1 Mb/s can be accomplished up to 37.4 km.

**Conclusion**

We compared two CV-QKD receiver architectures based on coherent heterodyne and intradyne detection methodology under the setting of a bandwidth-limited quantum detector. Even though it could be expected that the 2.5-fold symbol rate of the intradyne receiver (250 MHz vs. 100 MHz) would (over-)compensate for the additional 3-dB loss within the optical 90° mixer, especially when considering the lower excess-noise of the intradyne receiver ($\xi_{int}$ = 0.0153 vs. $\xi_{het}$ = 0.0182), the results prove the loss clearly being the performance-limiting factor. Therefore, the loss-conscious heterodyne CV receiver features a clear advantage for network applications with a link reach of >11.9 km (untrusted) and >15 km (trusted receiver scenario). Nonetheless, short-reach applications greatly benefit from the bandwidth-conscious intradyne CV receiver (9.2 – 5.4 dB higher SKR), especially under the scheme of periodic AES key renewal, where a complete exhaustion of generated key would require massive classical data capacities beyond 20 Pb/s.

**Acknowledgements**

This work was supported by the ERC under the EU Horizon-2020 programme (grant agreement No 804769). This work has further received funding from the EU Horizon 2020 programme under grant agreement No 820474.